\documentclass[prd,twocolumn,showpacs,reprint,preprintnumbers,nofootinbib,superscriptaddress]{revtex4-1}

\input{header}

\newcommand{\orcid}[1]{\begingroup
  \hypersetup{hidelinks}\href{https://orcid.org/#1}{\includegraphics[width=10pt]{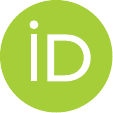}} \endgroup}

\begin{document}

\title{{\Large Discovering neutrino tridents at the Large Hadron Collider}}

\author{Wolfgang Altmannshofer~\orcid{0000-0003-1621-2561}}
\email{waltmann@ucsc.edu}
\affiliation{Department of Physics, University of California Santa Cruz, and
Santa Cruz Institute for Particle Physics, 1156 High St., Santa Cruz, CA 95064, USA\looseness=-1}

\author{Toni M\"akel\"a \orcid{0000-0002-1723-4028}\,}
\email{tmakela@uci.edu}
\affiliation{National Centre for Nuclear Research, Pasteura 7, Warsaw, 02-093, Poland}
\affiliation{Department of Physics and Astronomy, University of California, Irvine, CA 92697 USA\looseness=-1}

\author{Subir Sarkar
~\orcid{0000-0002-3542-858X}}
\email{subir.sarkar@physics.ox.ac.uk}
\affiliation{Rudolf Peierls Centre for Theoretical Physics, University of Oxford, Parks Road, Oxford OX1 3PU, United Kingdom\looseness=-1}

\author{Sebastian Trojanowski~\orcid{0000-0003-2677-0364}}
\email{sebastian.trojanowski@ncbj.gov.pl}
\affiliation{National Centre for Nuclear Research, Pasteura 7, Warsaw, 02-093, Poland}

\author{Keping Xie~\orcid{0000-0003-4261-3393}}
\email{xiekepi1@msu.edu}
\affiliation{Department of Physics and Astronomy, Michigan State University, East Lansing, MI 48824, USA}

\author{Bei Zhou \orcid{0000-0003-1600-8835}\,}
\email{beizhou@fnal.gov}
\affiliation{Theoretical Physics Department, Fermi National Accelerator Laboratory, Batavia, Illinois 60510, USA}
\affiliation{Kavli Institute for Cosmological Physics, University of Chicago, Chicago, Illinois 60637, USA\looseness=-1}

\preprint{FERMILAB-PUB-24-0294-T, MSUHEP-24-007}

\begin{abstract}
Neutrino trident production of di-lepton pairs is well recognized as a sensitive probe of both electroweak physics and physics beyond the Standard Model. Although a rare process, it could be significantly boosted by such new physics, and it also allows the electroweak theory to be tested in a new regime. We demonstrate that the forward neutrino physics program at the Large Hadron Collider offers a promising opportunity to measure for the first time, dimuon neutrino tridents with a statistical significance exceeding $5\sigma$, improving on the previous claims at the $\sim 3\sigma$ level by the CHARM-II and CCFR collaborations while accounting for additional backgrounds later identified by the NuTeV collaboration. We present predictions for various proposed experiments and outline a specific experimental strategy to identify the signal and mitigate backgrounds, based on ``reverse tracking'' dimuon pairs in the FASER$\nu$2 detector. We also discuss prospects for constraining beyond Standard Model contributions to neutrino trident rates at high energies.
\end{abstract}

\maketitle 
\clearpage

\section{Introduction} 
\label{sec:introduction}

The success of the Standard Model (SM) has been firmly established in particle physics over the last half century via numerous observations of new elementary processes. Intriguingly, several such pivotal studies concern neutrino interactions. These have long provided essential consistency tests of the SM, and also the first evidence of new physics beyond the Standard Model (BSM), namely neutrino mass, which motivates further attempts to measure very rare events involving neutrinos.

A good example of such a process is ``neutrino trident scattering'' off a nucleus, $\nu N\to\nu^{(\prime)} \ell^-\ell^{(\prime)+} N$, where $\ell$ and $\ell^{(\prime)}$ are charged leptons of the same or different flavor, and lepton number is conserved taking into account the incident and outgoing neutrinos, $\nu$ and $\nu^{(\prime)}$~\cite{Kozhushner1961, Shabalin1963, Czyz:1964zz, Lovseth:1971vv, Fujikawa:1971nx, Koike:1971tu, Koike:1971vg, Brown:1972vne, Brown:1973ih, Belusevic:1987cw, Altmannshofer:2014pba, Magill:2016hgc, Ge:2017poy, Ballett:2018uuc, Altmannshofer:2019zhy, Gauld:2019pgt, Zhou:2019vxt, Zhou:2019frk}. It is an attractive test of the electroweak theory as it is affected by both $Z$- and $W$-mediated contributions, depending on the final state. It has also been recognized as a sensitive probe of BSM physics~\cite{Altmannshofer:2014cfa, Altmannshofer:2014pba}. 

Measuring neutrino tridents is however difficult. To date, only a few experiments have reported results and only for events with a dimuon pair in the final state, $\nu N\to\nu \mu^+\mu^- N$. The CHARM-II~\cite{CHARM-II:1990dvf} collaboration used a glass target to study interactions of neutrinos with energies of order $E_\nu\sim 20~\textrm{GeV}$, while subsequent measurements in CCFR~\cite{CCFR:1991lpl} and NuTeV~\cite{NuTeV:1999wlw} detectors employed an iron target and more energetic neutrinos with $E_\nu\sim 160~\textrm{GeV}$. The first two experiments found evidence for such interactions at only the $\sim3\sigma$ level, while for the third, new backgrounds that had been previously overlooked have come to light. \textit{Thus, presently, no conclusive signal of trident events can be claimed.} This has motivated new proposals to measure neutrino tridents at forthcoming experiments --- using both accelerators and neutrino telescopes~\cite{Altmannshofer:2014pba,Magill:2016hgc,Ge:2017poy,Ballett:2018uuc,Altmannshofer:2019zhy,Zhou:2019vxt,Zhou:2019frk,Sarkar:2023dvr}. In particular, future measurements at DUNE could detect the dimuon trident signal with significance between $2$ and $4\sigma$~\cite{Altmannshofer:2019zhy}.

We propose to perform the first such measurement at TeV energies by utilizing the neutrinos produced in the very forward kinematic region at the collision points of the Large Hadron Collider (LHC). This provides a naturally well-collimated flux of neutrinos whose interactions can be studied in compact, well-instrumented detectors with high-$Z$ nuclear target material. Both these characteristics and the increased energy with respect to past searches allow maximization of the neutrino trident scattering rates and benefit from improved background rejection.\footnote{The trident cross sections are computed using a modified version of the code~\cite{Altmannshofer:2019zhy}, which is publicly available at \href{https://github.com/makelat/forward-nu-flux-fit}{https://github.com/makelat/forward-nu-flux-fit}, together with the software developed for the present study.} Importantly, the measurement at the LHC will not be dominated by a resonance or on-shell $W$-boson production, which is relevant at even higher energies~\cite{Seckel:1997kk,Alikhanov:2015kla, Zhou:2019vxt,Zhou:2019frk,Xie:2023qbn}. This allows for probing interference between charged current (CC) and neutral current (NC) contributions in the SM, as well as studying potential BSM effects.

These advantages situate the LHC, equipped with a dedicated forward neutrino detector, as an ideal facility for neutrino trident measurements in the near future. We show that even a modest $\sim 10~\textrm{ton}$ detector sensitive to forward neutrinos will detect tens of dimuon trident events during the High-Luminosity LHC (HL-LHC) era. This would extend the existing LHC neutrino physics program of the FASER/FASER$\nu$~\cite{FASER:2019dxq,FASER:2020gpr,FASER:2022hcn} and SND@LHC~\cite{SHiP:2020sos,SNDLHC:2022ihg} detectors. We present predictions for $\mu^+\mu^-$ and other dilepton final states, including the rare ditau production, and discuss specific prospects for the proposed Forward Physics Facility (FPF)~\cite{Anchordoqui:2021ghd, Feng:2022inv}.

This paper is organized as follows. Neutrino trident interactions are introduced in~\cref{sec:tridents}, and possibilities for detecting them at the LHC are discussed in~\cref{sec:LHCtridents}, along with the expected numbers of signal events at various existing and proposed detectors. Then, the remainder of this paper focuses on a single detector, FASER$\nu$2. \cref{sec:FASERnu2} details the expected signal, and \cref{sec:BG} discusses the background and discovery prospects for trident events. \Cref{sec:BSM} discusses how these observations can be used to constrain muonic four-Fermi interactions. Our conclusions are presented in~\cref{sec:conclusions} and
further details about the form factors used in computing the trident signal cross sections are given in~\cref{sec:formfactors}.

\section{Neutrino tridents} 
\label{sec:tridents}

\begin{figure}[t]
\includegraphics[width=0.22\textwidth]{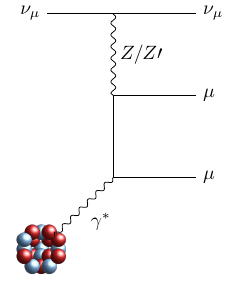}
\includegraphics[width=0.22\textwidth]{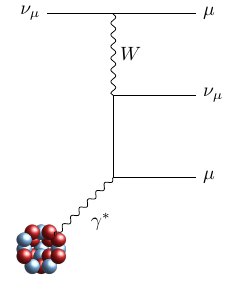}
\caption{Illustrative Feynman diagrams for trident production of a dimuon pair in interactions of incident muon neutrinos (see Fig.~5 of Ref.~\cite{Zhou:2019vxt} for all five diagrams). Both neutral and charged current contributions in the SM are shown; the former can have new physics contributions from e.g. the exchange of a $Z^\prime$ boson.}
\label{fig:diagrams}
\end{figure}

\begin{figure*}
\includegraphics[trim={13mm 0mm 0mm 0mm},clip,width=0.75\textwidth]{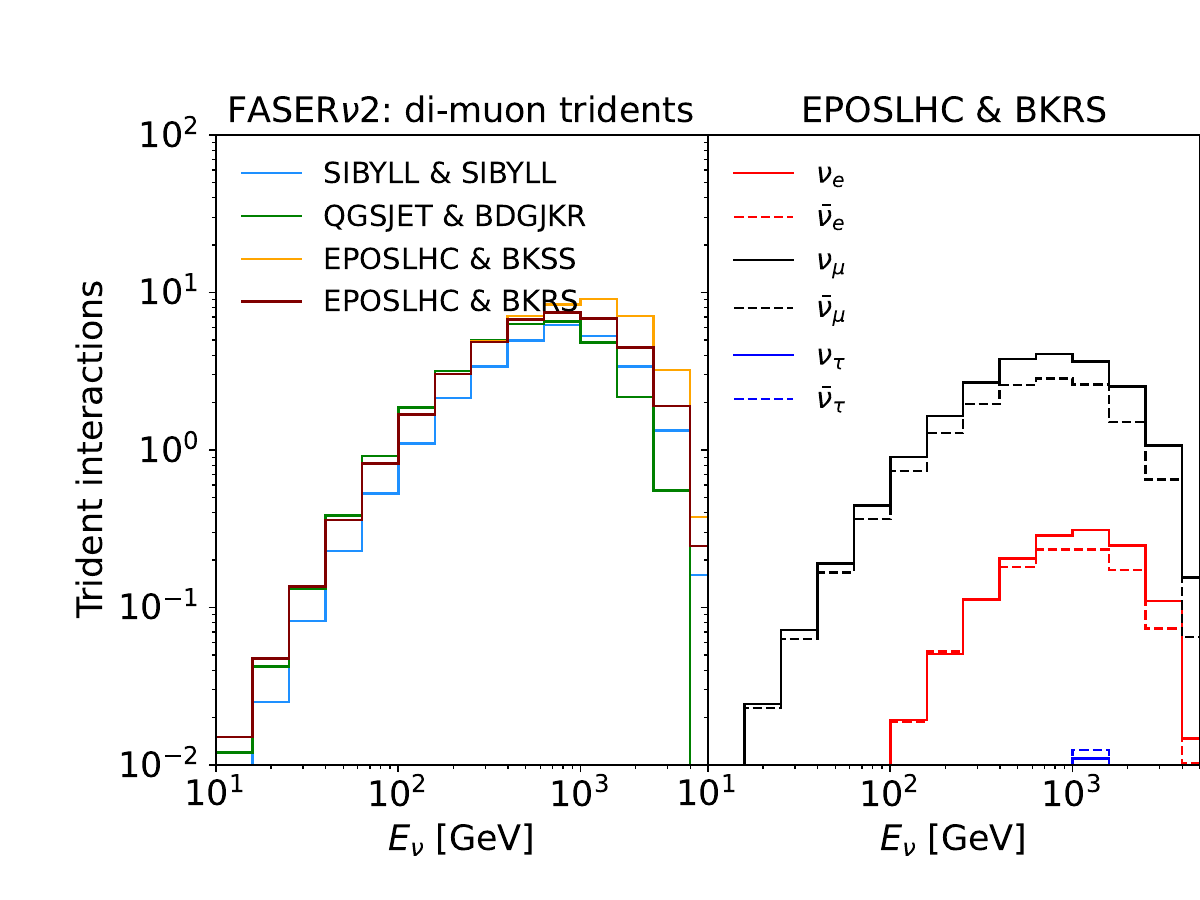}
\caption{\textit{Left:} The sum of trident dimuon signal events originating from all neutrino and antineutrino flavors is shown, using various neutrino flux predictions described in the text. 
\textit{Right:} The individual contributions of electron, muon, and tau neutrinos and antineutrinos to the signal, using the EPOSLHC\&BKRS neutrino flux prediction.}
\label{fig:trident_generators_contributions}
\end{figure*}

Neutrino trident interaction is the process in which a charged lepton pair is produced through neutrino scattering in a nuclear electric field. Both CC and NC  interactions can be relevant for the process depending on the charged-lepton final state. The left panel of \cref{fig:diagrams}  shows typical Feynman diagrams for dimuon production. Although the cross section is suppressed in the SM (compared to, e.g.,  $e^\pm\mu^\mp$ production) due to destructive interference between the dominant $W$- and subdominant $Z$-mediated contributions~\cite{Belusevic:1987cw}, it remains the most relevant experimentally, given the ease of identifying and measuring both outgoing muons. The relevant interaction rate can also increase if there is a light BSM vector mediator $Z^\prime$ with $m_{Z^\prime}\ll m_Z$~\cite{Altmannshofer:2014cfa,Altmannshofer:2014pba}, as is also indicated.

The neutrino trident cross section can be estimated using the equivalent photon approximation~\cite{vonWeizsacker:1934nji,Williams:1934ad}, in which the subprocess of the neutrino interaction with a real photon is first calculated and then convolved with the virtual photon distribution in the nuclear electric field~\cite{Belusevic:1987cw,Altmannshofer:2014pba,Magill:2016hgc}. However, in the most general case, this was found to introduce sizeable uncertainties that can only be reduced with a full calculation of the $2\to 4$ scattering process~\cite{Ballett:2018uuc, Altmannshofer:2019zhy, Zhou:2019vxt, Zhou:2019frk}. In the following, we employ the detailed calculation of Ref.~\cite{Altmannshofer:2019zhy}.

The dominant trident production mode in neutrino interactions with a nucleus is coherent scattering in which the nucleus stays intact. The corresponding cross section is thus $Z^2$-enhanced, where $Z$ is the atomic number of the nucleus. We refer to \cref{sec:formfactors} for the discussion of nuclear form factor parameterizations used in our study. The coherent scattering cross section calculation suffers from small theoretical uncertainties from modeling these form factors, as well as higher-order QED effects, and ambiguity in the weak mixing angle predictions. However, these uncertainties are estimated to be only a few percent, so will not affect our conclusions~\cite{Altmannshofer:2019zhy}.

Additionally, we include diffractive contributions from scattering off individual nucleons. The production rate is now only $Z$-enhanced and dominated by scattering off protons. Depending on the target material and the charged-lepton-pair final state, this typically contributes no more than a few $\%$ to the total signal for the heavy target nuclei of our interest. Only in some cases do diffractive scatterings become more important and contribute up to about $20-30\%$ of the total cross-section. This effect is strengthened for nuclear targets with a smaller $Z$ and for heavier final-state charged leptons. The production of tau leptons, especially, requires a larger momentum exchange that favors more substantial diffractive contributions. However, although we report the corresponding signal rates below, we stress that processes involving tau leptons in the final state are phase-space suppressed with respect to other neutrino trident production modes and thus remain challenging to probe experimentally.

In modeling diffractive trident production, we introduce additional cuts on the energy of the final-state protons to ensure that they remain below the detection threshold in the detectors of interest. Consequently, such contributions are suppressed further, and diffractive processes contribute no more than $\sim 10\%$ of the coherent-like signal that we focus on. 
Diffractive trident production may be even more strongly suppressed if energy dissipation due to final state interactions (FSI) of the proton inside the nucleus is considered. We do not model neutrino trident production in deep inelastic scattering (DIS), as this has substantial backgrounds in the experiments discussed below. We also do not consider final-state radiation from the final-state charged leptons, which is not important for $E_\nu \lesssim 1$ TeV~\cite{Plestid:2024bva}.
We conclude that the signal of our interest is mainly associated with coherent neutrino scattering, with only a modest diffractive contribution. The uncertainties in estimating the latter thus play little role in our analysis.

\section{Neutrino Tridents at the LHC}
\label{sec:LHCtridents}

Besides its main physics goals, the LHC is also a neutrino factory with most of the high-energy neutrinos produced in the very forward kinematic region ~\cite{DeRujula:1984pg,Winter:1990ry,Vannucci:1993ud,DeRujula:1992sn,Park:2011gh,Feng:2017uoz}. This observation has led to a dedicated neutrino physics program in the Run 3 data-taking period, and the first observations of neutrino interactions at the LHC have recently been reported~\cite{FASER:2021mtu,FASER:2023zcr,SNDLHC:2023pun,FASER:2024hoe}. The FASER and SND@LHC experiments are foreseen to continue taking data beyond the current period and potentially expand their scope. In addition, a significant extension of this program has been proposed for the future HL-LHC era, in which the FPF would host several experiments targeting neutrino and QCD physics with implications for astroparticle and BSM searches~\cite{Anchordoqui:2021ghd, Feng:2022inv, Amoroso:2022eow}. Importantly, the LHC neutrino physics program supports and complements the astroparticle high- and ultra-high energy neutrino program~\cite{Ackermann:2022rqc}.

The proposed neutrino trident measurement at the LHC is thus well-grounded in the ongoing and proposed experimental efforts. The forward neutrino spectrum at the LHC peaks at  several hundred GeV~\cite{Kling:2021gos, FASER:2024ykc}. At these energies, the predicted SM neutrino trident cross section corresponds to $\mathcal{O}(0.01\%)$ of the total neutrino deep-inelastic scattering (DIS) cross section~\cite{Altmannshofer:2019zhy, Zhou:2019vxt}, resulting in tens of expected events for a $\mathcal{O}(10~\textrm{ton})$ detector operating in the HL-LHC era. While neutrinos of any flavor can produce neutrino tridents, the dominant contribution comes from muon neutrinos $\nu_\mu$ and antineutrinos $\bar{\nu}_\mu$ since their flux is the largest.

Notably, the incident neutrino flux is subject to uncertainties that currently range from tens of percent to even an order of magnitude, depending on the neutrino flavor and energy. The main uncertainty is the modeling of forward parent hadron spectra produced in $pp$ collisions at the LHC. However, these uncertainties can in principle be reduced to even sub-percent level thanks to neutrino DIS data to be gathered during the current Run 3 and future HL-LHC era~\cite{Kling:2023tgr}. The predicted rate of the neutrino trident process will then also be known with significantly higher precision. 

\cref{fig:trident_generators_contributions}, left panel shows our predicted neutrino trident interaction spectra for the dimuon final state. These apply to the proposed FASER$\nu$2 experiment near the ATLAS interaction point and are obtained using different available predictions for the spectra of light mesons (i.e., pions and kaons) and charmed hadrons. We model them separately, as indicated by two names in the labels in the plot, before and after the ampersand (\&) sign, for light and charmed hadrons, respectively. The generation of light mesons is simulated using \texttt{EPOS-LHC}~\cite{Pierog:2013ria}, \texttt{QGSJET~II-04}~\cite{Ostapchenko:2010vb}, or \texttt{SIBYLL~2.3d}~\cite{Ahn:2009wx,Ahn:2011wt,Riehn:2015oba,Fedynitch:2018cbl} generators as implemented in the \texttt{CRMC} package~\cite{CRMC}. To describe charm hadron production, we utilize the predictions using SIBYLL, but also perturbative calculations referred to as \texttt{BKRS}~\cite{Buonocore:2023kna}, \texttt{BKSS} $k_T$~\cite{Bhattacharya:2023zei} and \texttt{BDGJKR}~\cite{Bai:2020ukz,Bai:2021ira,Bai:2022xad} --- see also Ref.~\cite{Kling:2023tgr} for a summary and further references. Given the parent hadron spectra, the neutrino fluxes are obtained via the fast neutrino flux simulation, which models the propagation of hadrons through the LHC beam pipe and magnetic fields, and their decay into neutrinos~\cite{Kling:2021gos}.

The neutrino trident dimuon production rates range from $27$ expected coherent events for the SIBYLL+SIBYLL case, to $50$ such events predicted with the EPOSLHC+BKSS hadron spectra. These numbers correspond to $3~\textrm{ab}^{-1}$ of integrated luminosity in $pp$ collisions at the HL-LHC era. The difference between the predictions is mainly associated with the high-energy part of the muon (anti)neutrino spectrum, which is affected by their production in charmed hadron decays. We employ an intermediate EPOSLHC+BKRS prediction as a baseline for further study. After adding both the coherent and diffractive contributions, it predicts $40$ $\mu^+\mu^-$ trident events in FASER$\nu$2. In the right panel of \cref{fig:trident_generators_contributions}, we also present relative contributions of the incident electron, muon, and tau (anti)neutrinos to the expected dimuon neutrino trident signal rate. 

\begin{table*}
\centering
\begin{tabular}{c|c|c|c|c|c|c||c|c|c|c|c}
\hline
\hline
Name & Mass & Target & On(Off)- & $L^{-1}_{\textrm{int,TeV$\nu\to\mu\mu$}}$ & \multicolumn{7}{c}{Neutrino Tridents, $\nu N \to \nu N' \ell^+ \ell^-$}\\
 & [tons] & nucleus & -Axis & $\times 10^{17}$ [$\textrm{cm}^{-1}$] & $\mu^+\mu^-$ & $\mu^+\mu^-_{f_s=0.5}$ & $e^+e^-$ & $\tau^+\tau^-$ & $e^\pm\mu^\mp$ & $e^\pm\tau^\mp$ & $\mu^\pm\tau^\mp$ \\
\hline 
\multicolumn{12}{c}{Run 3 ($150~\textrm{fb}^{-1}$)} \\ 
\hline
FASER$\nu$ & 1.1 & W & On & 252 & 0.22 & 0.54 & 0.24 & 0.0029 & 0.83 & 0.035 & 0.060\\
SND@LHC & 0.83 & W & Off & 252 & 0.024 & 0.06 & 0.03 & 0.0002 & 0.10 & 0.004 & 0.004\\
\hline
\multicolumn{12}{c}{HL-LHC ($3~\textrm{ab}^{-1}$)} \\ 
\hline
FASER$\nu$2 & 20 & W & On & 252 & 40 & 97 & 44 & 0.51 & 150 & 6.3 & 10\\
AdvSND@LHC (Far) & 5 & W & Off & 252 & 2.2 & 5.3 & 2.7 & 0.02 & 9.0 & 0.3 & 0.4\\
FLArE & 10 & LAr & On & 8.56 & 4.5 & 11 & 4.5 & 0.07 & 16 & 0.7 & 1.2\\
FLArE-100 & 100 & LAr & On & 8.56 & 26 & 63 & 27 & 0.37 & 91 & 4.1 & 6.8\\
NuTeV-like (Fe) & 95 & Fe & On & 65.4 & 21 & 52 & 22 & 0.29 & 76 & 3.4 & 5.5\\
NuTeV-like (Pb) & 135 & Pb & On & 154 & 48 & 116 & 57 & 0.45 & 190 & 7.0 & 10\\
\hline
\hline
\end{tabular}
\caption{The expected number of neutrino trident events in both the neutrino detectors active during current LHC Run 3, as well as detectors proposed for the future HL-LHC era.  The numbers increase with target atomic mass number, inverse interaction length $L^{-1}_{\textrm{int,TeV$\nu\to\mu\mu$}}$, and total detector mass, with most interactions occurring along the line of sight from the LHC beam axis. For completeness, projections are shown for all charged lepton final states. For the $\mu^+\mu^-$ final state, an alternative calculation is also performed, in which a fraction $f_s=0.5$ of neutrino parent pions are translated into kaons, motivated by the enhanced strangeness scenario discussed in Ref.~\cite{Anchordoqui:2022fpn}.}
\label{tab:events}
\end{table*}

In \cref{tab:events} we provide the estimated event rates for various ongoing and proposed experiments. As can be seen, the ongoing neutrino measurements in FASER$\nu$ and SND@LHC are not expected to have sensitivity to measure neutrino trident production during LHC Run 3. However, the proposed future experiments at the FPF~\cite{Feng:2022inv} and beyond would be able to detect a few tens of such events and study them for different target materials. These include the FASER$\nu$2 emulsion detector which we discuss in greater detail in \cref{sec:FASERnu2,sec:BG}. The Table includes other proposed detectors such as the liquid argon (LAr) time-projection chamber experiment FLArE~\cite{Batell:2021blf}, the successor of the SND@LHC experiment dubbed Advanced SND@LHC, and a NuTeV-like $\mathcal{O}(100~\textrm{ton})$ detector in the forward region of the LHC~\cite{NuTeVFelix}. For the last case, we assume a detector of size $3~\textrm{m}\times 3~\textrm{m}\times 4~\textrm{m}$, with either iron or lead target material. \cref{tab:events} contains additional information on the target mass and nucleus for all the experiments, in particular if the detector is positioned on or off the proton beam axis. This is important as the highest energy neutrinos are collimated along the beam axis, hence detectors placed off-axis lose sensitivity.

In order to facilitate comparison between different detectors, we additionally provide in \cref{tab:events}  the inverse interaction length for the dimuon coherent trident production process for an incident muon neutrino with $E_\nu = 1~\textrm{TeV}$. We denote this by $L^{-1}_{\textrm{int,TeV$\nu\to\mu\mu$}} = (\rho_\mathrm{T}/m_\mathrm{T})\times \sigma_{\nu\to\mu\mu}$, where $\rho_\mathrm{T}$ and $m_\mathrm{T}$ are the target density and the target nucleus mass, respectively, and $\sigma_{\nu\to\mu\mu}$ is the cross section for the aforementioned process at 1 TeV. The larger the inverse interaction length, the larger interaction probability, as $P_{\textrm{int}} = L_{\textrm{det}}\times L^{-1}_{\textrm{int,TeV$\nu\to\mu\mu$}}$, where $L_{\textrm{det}}$ is the detector length. Most straightforward is the comparison between the NuTeV-like iron and lead detectors, which differ only in the target material. The inverse interaction length of Pb is about a factor of two larger than that of Fe, which translates into a similar difference in the reported number of $\mu^+\mu^-$ events. Note that the event numbers in the Table have been obtained after convoluting with the full neutrino spectrum. 

The tungsten (W) target material is characterized by a larger value of $L^{-1}_{\textrm{int,TeV$\nu\to\mu\mu$}}$ than lead (Pb). This difference, and also the different geometry, partially compensate for the expected \textsl{a priori} larger difference between a $135$-ton NuTeV-like Pb detector and $20$-ton tungsten FASER$\nu$2. The latter (geometry) effect is due to an increased tungsten density and a different shape of the proposed detector, which is more focused around the beam collision axis with a transverse size of only $40~\textrm{cm}\times 40~\textrm{cm}$. Consequently, FASER$\nu$2 captures most of the high-energy neutrino flux, and the more extensive transverse size of the NuTeV-like detector adds less than anticipated based on its size alone. Therefore, the expected number of dimuon trident events is similar for both detectors. The effect of the neutrino flux collimation can also be seen when FLArE and FLArE-100 predictions are compared; even though the latter has an order of magnitude larger target mass, the event rate is not proportionally enhanced due to the increased transverse size of FLArE-100. For this reason and the difference in $L^{-1}_{\textrm{int,TeV$\nu\to\mu\mu$}}$ between LAr and tungsten, even the $100$-ton FLArE detector has a lower expected dimuon trident event rate than FASER$\nu$2.

If any non-standard effects change the incident neutrino spectrum, they will also affect the predicted dimuon trident production rate. In particular, it has been proposed~\cite{Anchordoqui:2022fpn} that a significantly enhanced strangeness production at large rapidities in $pp$ collisions could explain the persistent \textsl{cosmic-ray muon puzzle}~\cite{Albrecht:2021cxw} while respecting constraints from measurements at lower energies and in central LHC detectors. The corresponding predicted change in the neutrino spectra will substantially increase the number of expected neutrino trident events. We provide the corresponding expectations in \cref{tab:events}  denoted by the label $f_s = 0.5$, where $f_s$ describes the pion to kaon swapping probability. See Refs.~\cite{Anchordoqui:2022fpn,Kling:2023tgr} for further discussion regarding the impact on the forward LHC data.

We also show in \cref{tab:events} the predicted trident event rates for dielectron, ditau, and mixed final states with two charged leptons of a different flavor. As can be seen, the predicted number of $e^\pm\mu^\mp$ production events is the largest being of order $\mathcal{O}(100)$ for the FASER$\nu$2 and the NuTeV-like detectors. The relevant production process in this case is the dominant charged-current contribution, and there is no destructive interference with the $Z$-mediated diagram. This final state, however, is difficult to study experimentally. A similar statement is valid for the interactions producing a tau lepton(s). In this case, the aforementioned scaling between different experiments breaks down further due to the diffractive contribution. We find that up to $\mathcal{O}(10)$ mu-tau events can be expected in the detectors considered.

\section{Neutrino trident signal in FASER$\nu$2}
\label{sec:FASERnu2}

Detecting neutrino trident events requires a dedicated search strategy. In this section, we discuss experimental cuts that optimise the capabilities of the FASER$\nu$2 detector proposed for the FPF to study dimuon trident production. We focus first on signal selection and then discuss backgrounds in \cref{sec:BG}.\footnote{See also Ref.~\cite{Francener:2024wul} for a recent study, which also notes that the observation of the neutrino trident process is, in principle, feasible at the Forward Physics Facility, but does not outline an experimental strategy to identify signal events or study the expected backgrounds, as we have done in detail, in order to establish the necessary criteria for a discovery.}

\begin{figure*}
\centering
\includegraphics[trim={0mm 0mm 0mm 0mm},clip,width=1.0\textwidth]{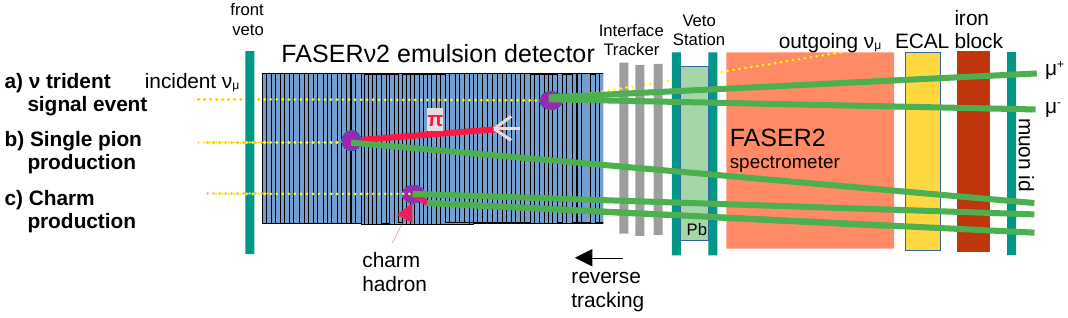}
\caption{
Illustration of the neutrino trident production signature and the two main background processes (single pion production and charm production) in high-energy neutrino interactions in the FASER$\nu$2 detector. In the trident signal, two outgoing muons (shown in green) leave long tracks in the emulsion in FASER$\nu$2 and travel through the Interface Tracker (IF), veto station, spectrometer, electromagnetic calorimeter, and iron block to be identified as muons. Background events contain at least one charged hadron track (shown in red). This could be long (for pions and kaons) and typically ends with the hadronic interaction vertex in tungsten, lead, or iron block. It could also be short and produce the second muon mimicking the trident signature. Our event analysis and background rejection rely on reverse tracking between the IF and the emulsion. See the text for more details.\label{fig:schematic}}
\end{figure*}

We consider first coherent-like neutrino trident signal events that produce only two charged tracks in the final state, i.e. a pair of muons. While other charged lepton pair final states could also be detected in FASER$\nu$2, mitigating backgrounds would be challenging in the absence of long muon tracks. This is especially relevant for electrons or positrons in the final state that generate electromagnetic showers in the emulsion. The discussion below may also apply to the interesting case of the $\mu^\pm\tau^\mp$ final state, which has never been observed. This is due to possible tau lepton decays into muons with branching fraction $\textrm{BR}(\tau^-\to\mu^-\bar{\nu}_\mu\nu_\tau) = 0.1739$~\cite{ParticleDataGroup:2022pth}. The resulting event will resemble dimuon trident production with an intermediate short tau track, which could also be resolved for $E_\tau\gtrsim \textrm{few tens of GeV}$ thanks to the excellent spatial resolution of the emulsion detector. However, the expected number of events in FASER$\nu$2, cf. \cref{tab:events}, is not sufficient to guarantee the detectability of such events after cuts, and a larger detector would be needed for this purpose.\footnote{See Ref.~\cite{Bigaran:2024zxk} concerning prospects for measuring tridents with taus in the final state at DUNE.} We focus therefore on the most promising $\mu^+\mu^-$ final state below.

\cref{fig:schematic} shows schematically the FASER$\nu$2 detector along with the signal event signature and leading backgrounds. This detector will employ emulsion detectors and tungsten target material~\cite{Feng:2022inv}. We assume that the 2$~\textrm{mm}$ thick tungsten plates are interleaved with emulsion layers and the interaction vertices in the tungsten can be reconstructed based on the tracks of charged particles in the emulsion. The events in the emulsion detector are not time-stamped, however this can be mitigated by employing the interface tracker (IF) station placed downstream of the detector. The muons produced in neutrino interactions in the emulsion layers propagate to the IF station, after which their momenta are measured in the FASER2 spectrometer, cf. the discussion for the current FASER detector~\cite{FASER:2020gpr}. The trident processes also have a missing energy component due to the undetected final state neutrino.

The role of the IF station is to allow for connecting the neutrino interaction vertex in the emulsion and the outgoing long muon track with the muon momentum measurement in the spectrometer. This timing information is necessary to facilitate muon charge identification and improve the accuracy of the muon energy measurement. To this end, when neutrino events are studied, first, a vertex is identified with at least five charged tracks in the final state. The long muon track in the emulsion is matched with the IF event. 

For neutrino trident events, the interaction vertex has only two charged tracks. To facilitate signal identification, we propose to use a novel \textsl{reverse tracking} strategy, i.e. the triggering of the events will be based on the spectrometer and IF station measurement of two time-coincident muons and no simultaneous front veto layer activation. The latter cut is introduced to reject muons coming from the ATLAS interaction point. Neutrino-induced muons produced in tungsten and detected in the spectrometer will be reverse-matched with the tracks in the emulsion. To this end, a sample of tracks in the emulsion that are consistent with the candidate IF station tracks should first be identified to reduce the combinatorial complexity of the matching. The matched tracks will then be reverse-tracked in the emulsion to verify if they come from the common vertex that satisfies other cuts discussed below. Before the interaction vertex is analyzed in FASER$\nu$2, the predicted neutrino-induced dimuon background rate is of $\mathcal{O}(7\textrm{k})$, as discussed below. This is also the expected triggering rate of such events in the entire HL-LHC period, for which the reverse tracking analysis would be performed. This triggering strategy greatly simplifies the search, which otherwise would require searching for all vertices in the emulsion with two outgoing muons and no incoming tracks.

Our predicted rates correspond to events with the two muons in the final state properly identified, which is an essential part of the signature. The muon identification is based on their long tracks in the emulsion, the IF and spectrometer matching, and the final muon ID detector. The latter is placed downstream of the FASER2 spectrometer and the electromagnetic calorimeter (ECAL) shown in yellow in \cref{fig:schematic}. It is additionally shielded with an iron block to help disentangle charged pions from muons. In \cref{fig:schematic}, we also show the additional veto layer between the emulsion detector and spectrometer. It is interleaved with a lead plate with a thickness of more than a few radiation lengths, as indicated by the green rectangle marked Pb. This layer improves the rejection of events with photons or electrons produced in the final state that could occasionally exit the emulsion detector before showering. 

The reverse tracking signature requires both muons produced in the neutrino trident process to leave the emulsion detector without losing all their energy while traversing the tungsten layers. In the high-energy regime of our interest, the dominant muon energy loss mechanisms are ionization, radiative bremsstrahlung, direct $e^+e^-$ pair production, and photonuclear interactions~\cite{Groom:2001kq}. In order to consider this effect, we model the muon energy loss via their mean stopping power for tungsten~\cite{ParticleDataGroup:2022pth}. The longitudinal position of a neutrino interaction vertex in the emulsion detector is assumed to follow a uniform distribution. This position determines the number of tungsten layers crossed by the outgoing muons. We find that muons with energies of $E_\mu\gtrsim (20-30)~\textrm{GeV}$ typically reach the IF station and spectrometer. They can then be measured and time stamped. 

We note that muons with $E_\mu\lesssim \textrm{a few tens of GeV}$ may be deflected away by the strong magnetic field of the spectrometer before going through all the tracking stations, ECAL, and the iron block. This could affect their muon ID, although they are still expected to leave long charged tracks in the emulsion if they originate from well inside the detector. However, events originating from the last few tens of cm of the emulsion can be negatively affected by the spectrometer magnets and this should be taken into account, depending on the final design of the magnet. While high-energy muons with $E_\mu\gtrsim 100~\textrm{GeV}$ will be well measured and identified in FASER2, low-energy muons with $E_\mu\lesssim 10~\textrm{GeV}$ are often stopped in tungsten, thereby preventing the use of the reverse tracking strategy discussed above.

\begin{figure}
\centering
\includegraphics[width=0.4\textwidth]{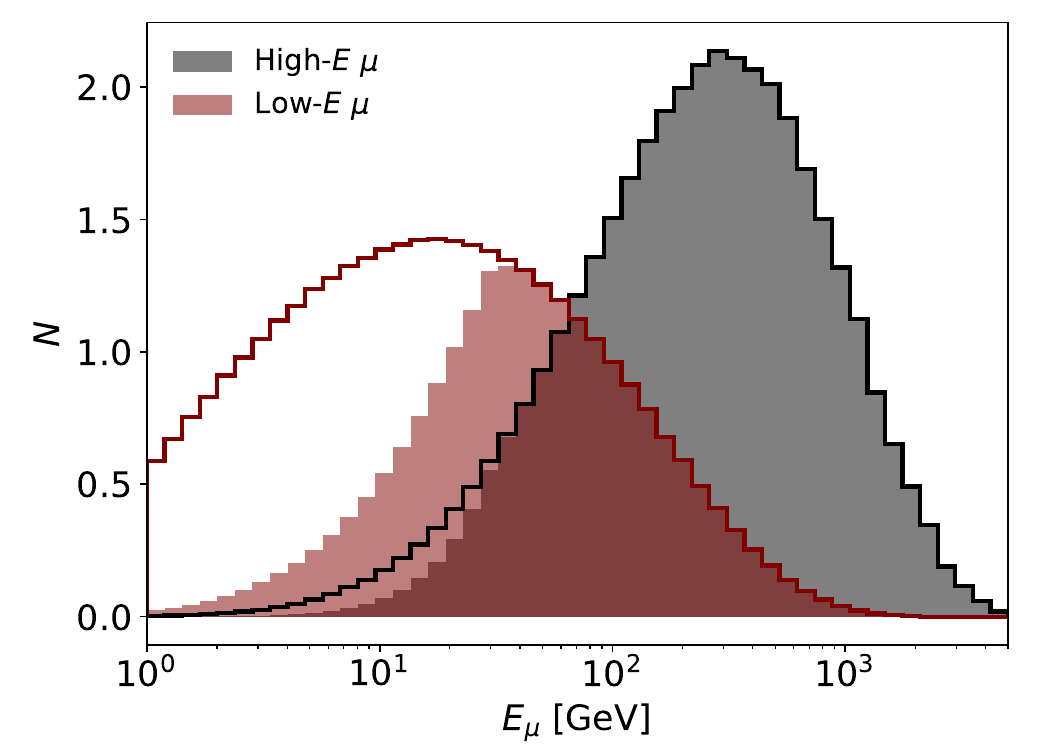}
\caption{Energy distributions of the lower and higher energy muons in the trident signal sample. The outlined histogram indicates the full spectrum without consideration of muon energy loss, while the filled regions indicate the muons expected to be within FASER$\nu$2 acceptance.
}
\label{fig:signal_E_mu_dist}
\end{figure}

\cref{fig:signal_E_mu_dist} illustrates the impact of the muon energy-loss cut on the outgoing muon spectra in FASER$\nu$2. In the plot, we show the energy distribution of both the final-state muons produced in the neutrino trident production process. As can be seen, typically, one of the outgoing muons is very energetic, with $E_\mu$ of order a few hundred GeV, and the other is softer with tens of GeV energy. The latter muon is then primarily subject to the efficiency cut, and we show the impact of such a cut. We find that about $50\%$ of the neutrino trident events in FASER$\nu$2 survive the muon-energy-loss cut and can be studied with the reverse tracking strategy.

In order to disentangle neutrino trident dimuon production events from the backgrounds discussed below, we introduce further cuts on the vertex identification and the angle between the two muons. In particular, we require that only two charged tracks with energy above the detectability threshold for FASER$\nu$2 emerge from the vertex. This threshold is set to $300~\textrm{MeV}$ in the following~\cite{FASER:2019dxq}. We also reject all events in which muons are not reconstructed as coming from a single vertex. We specifically require that the muon production vertices cannot be separated by more than $2~\textrm{mm}$ distance, corresponding to the assumed tungsten plate thickness. Eventually, we require the opening angle between the two muons to be no larger than $0.1~\textrm{rad}$. This relies on the perfect capability to measure the angle between charged tracks in the emulsion detector, which is expected to have sub-mrad precision for extended tracks~\cite{FASER:2019dxq}. We find that applying the above cuts suppresses the neutrino trident signal rate only slightly, and about $93\%$ of them survive.

We summarize below the cuts employed:
\begin{description}
\item[Muon ID \& neutral vertex] We reject events initiated by incident muons (charged vertices) or containing energetic charged pions or kaons in the final state.
\item[Muon-energy-loss] We apply the efficiency cut on both muons to reach the IF station and spectrometer, cf. \cref{fig:signal_E_mu_dist} and the discussion above.
\item[Multiplicity] We require only two charged tracks emerging from the neutrino interaction vertex.
\item[Displaced vertex] Both the final-state muons must have a common vertex, i.e., their tracks should start at most $2~\textrm{mm}$ apart from each other.
\item[Muon opening angle] The angle between the two outgoing muons should not exceed $0.1~\textrm{rad}$.
\end{description}

\section{Backgrounds in FASER$\nu$2\label{sec:BG}}

The FASER$\nu$2 detector in the FPF would be very well shielded from the ATLAS interaction point (IP), such that only the neutrinos and muons produced there can reach the detector. Both create backgrounds in the neutrino trident measurement. The high-energy neutrinos dominantly originate from decays of hadrons near the IP or in the beam pipe before the parent hadrons are deflected by LHC magnets or hit particle absorbers. The flux of these forward neutrinos is suppressed only at low energies, $E_\nu\sim 1~\textrm{GeV}$, in which case additional production modes in hadronic showers down the beam pipe become important~\cite{Kling:2021gos}. These, however, do not generate background events that can mimic the neutrino trident signal and pass the cuts. High-energy muons are also dominantly produced near the ATLAS IP. However, as the LHC magnets can deflect these, additional secondary muon production modes are also considered when modeling their flux and spectrum, namely proton collisions with the elements of the LHC infrastructure down the beam pipe~\cite{Feng:2022inv}. We employ the results of such simulations performed by the CERN STI group in our analysis below~\cite{sabate-gilarte:ipac2023-mopl018}. We now discuss how to suppress neutrino- and muon-induced backgrounds using the cuts above.

\subsection{Neutrino-induced backgrounds\label{sec:neutrinoinducedBG}}

Backgrounds related to neutrino interactions can be broadly categorized into two main types. In one case, direct non-trident muon pair production occurs; this is typically connected to an intermediate heavier hadron decaying into the muon. Alternatively, a muon and long-lived charged hadron (pion or kaon) can be present in the final state, where the latter mimics a muon in the detector. 

\begin{figure}[t]
\centering
\includegraphics[width=0.45\textwidth]{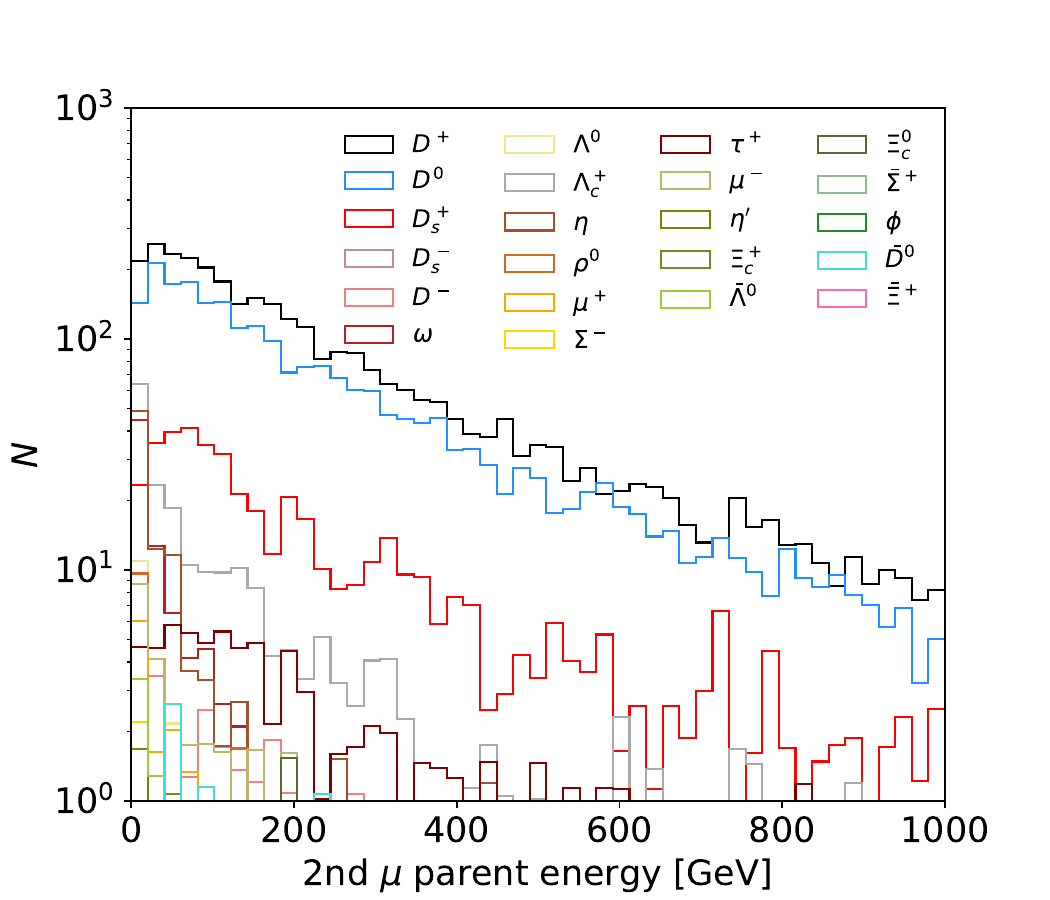}
\caption{Prompt dimuon backgrounds in the FASER$\nu$2 detector. The energy distribution of the parent hadrons of the 2nd muon (i.e., \emph{not} the one from the leptonic vertex in the neutrino interaction) is shown, for each hadron type. Only production channels that are expected to create more than one event are included.
}
\label{fig:bg_parent_hadrons}
\end{figure}

\medskip
\textbf{Prompt dimuon production (non-trident)}
The majority of the background to the dimuon neutrino trident process is due to charge current DIS (CCDIS) events involving charmed meson production. These mesons can decay quickly into muons, mimicking the signature of trident events. 
This background is simulated using {\tt PYTHIA8}~\cite{Sjostrand:2014zea}, following the procedure initially developed for a study focusing on the CCDIS dimuons~\cite{Zhou:2021xuh}. 
We estimate 4579 (1751) such interactions are originated by $\nu_\mu$ ($\overline{\nu}_\mu$) within the FASER$\nu$2 detector, with the muons predominantly produced in $D$ meson decays as shown in \cref{fig:bg_parent_hadrons}. Smaller contributions from hyperon production and decay, as well as other processes, are also shown as a function of the parent hadron energy. As this background is mostly inelastic, it can be reduced to a negligible level by the experimental cuts proposed in \cref{sec:FASERnu2} as follows.

\begin{figure}[t]
\centering
\includegraphics[width=0.4\textwidth]{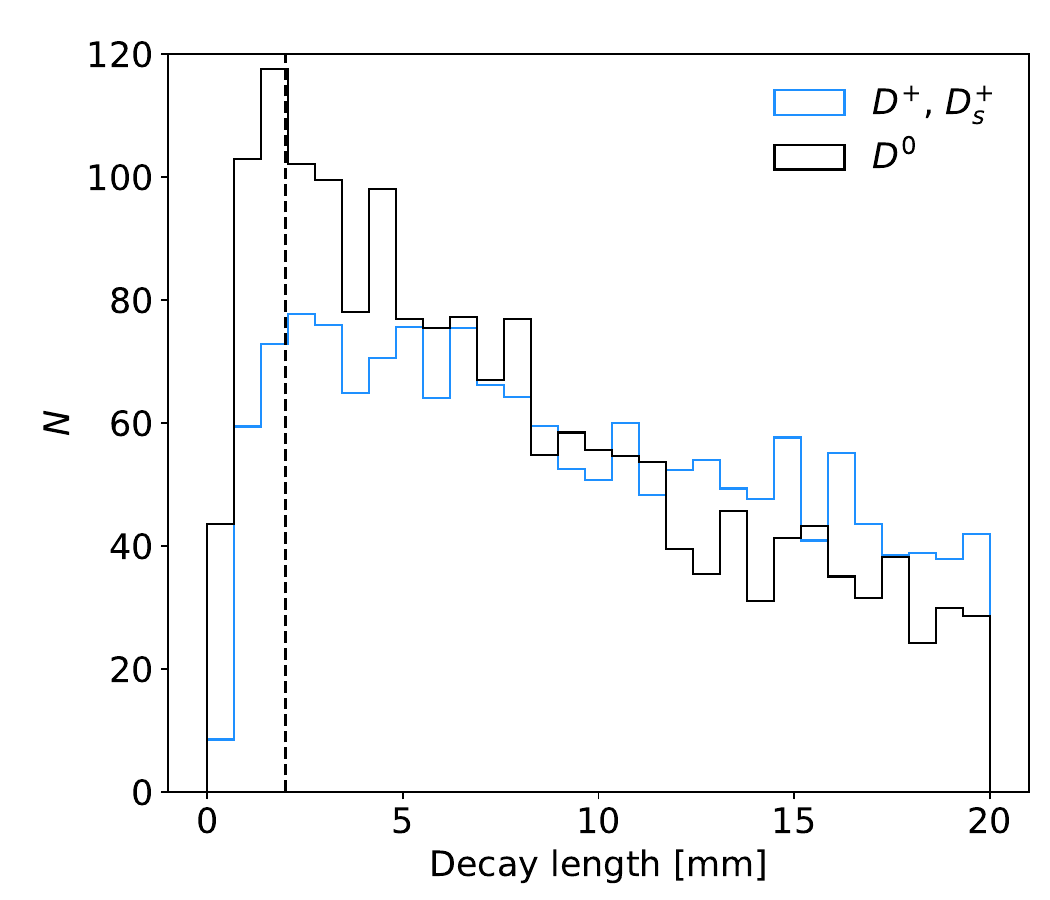}
\caption{The decay length distribution in the background sample obtained for a prompt dimuon production in the FASER$\nu$2 detector. The events to the right of the dashed line are excluded by the displaced vertex cut.
}
\label{fig:D_decay_lengths}
\end{figure}

The displaced vertex cut removes a significant portion of the $D^\pm$ and $D^\pm_s$ events, but is less effective against $D^0$ events as illustrated in \cref{fig:D_decay_lengths}, which shows the decay length distribution of charmed mesons produced in muon neutrino interactions in FASER$\nu$2. While $c\tau < \textrm{mm}$ for each of these mesons (where $\tau$ is the lifetime), they are typically produced with a boost factor $\gamma$ of  $\mathcal{O}(10)$ in high-energy neutrino scatterings relevant for our analysis. Therefore, their decay length is driven to larger values above the assumed $2~\textrm{mm}$ track reconstruction resolution indicated with a vertical dashed line in the plot. In particular, charged mesons can be identified based on this, as they leave tracks in the emulsion.

\begin{figure}[t]
\centering
\includegraphics[width=0.4\textwidth]{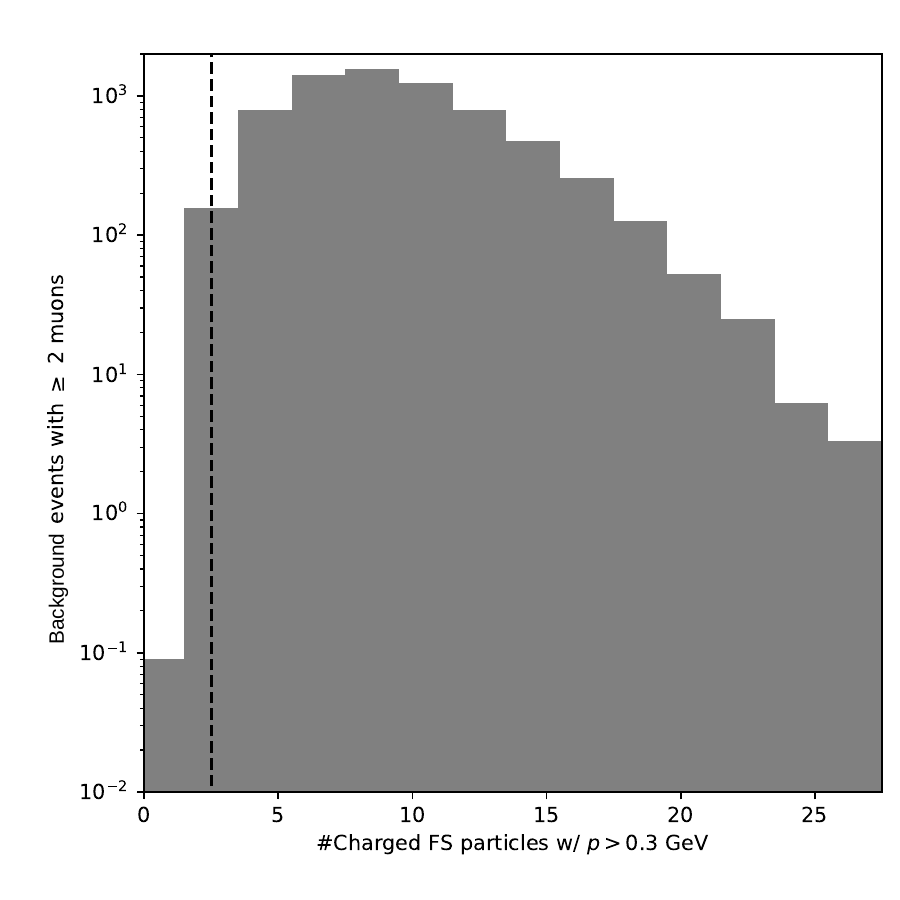}
\caption{The multiplicity of charged tracks in the background sample obtained for a prompt dimuon production in the FASER$\nu$2 detector. The dashed line illustrates that the bulk of the background is removed, and less than $10$ events remain after restricting to exactly two charged tracks.}
\label{fig:bg_convoluted_Nch}
\end{figure}

Note that although the $D^0$ does not leave a charged track, its daughter muons are necessarily accompanied by another charged track, thus violating the multiplicity cut. 
\cref{fig:bg_convoluted_Nch} shows that the multiplicity cut is highly effective against most DIS dimuon backgrounds. A related source of background events is $\nu_\tau$ CC scattering producing a $\tau + D$ meson, with the $\tau$ decaying as $\nu_\tau + \mu + \bar{\nu}_\mu$. This is, however, suppressed by the low rate of $\nu_\tau$ events and reduced further by the displaced vertex and multiplicity cuts akin to $D$ meson production via $\nu_\mu$-CC-DIS.

It is also possible that two or more muons are produced in NC interactions. However, unlike the signal events, these are not expected to have a high-energy muon inheriting most of the neutrino momentum and accompanied by a softer near-parallel muon produced in the Coulomb field. Hence, such events with multiple muons originating from hadronic decays are effectively removed by the charged track multiplicity and muon opening angle cuts.

\begin{figure}[t]
\centering
\includegraphics[width=0.4\textwidth]{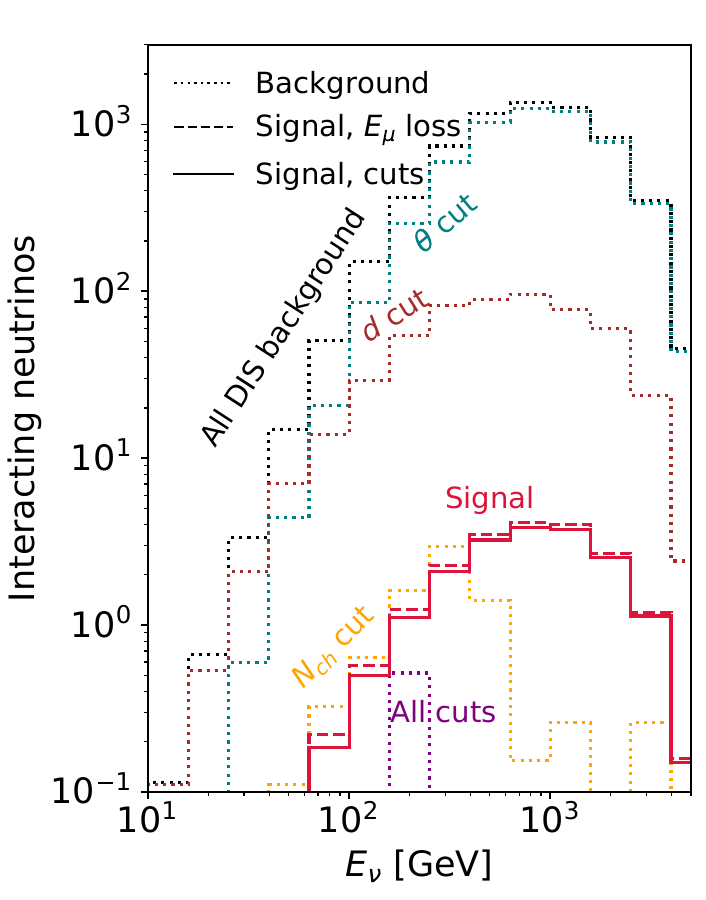}
\caption{The effect of various cuts on the backgrounds, both individually as well as in combination. The dashed red histogram corresponds to the dimuon trident signal rate after the muon energy loss is considered, i.e., only events that permit reverse tracking of both muons are included. The solid red histogram shows the effect of the applied cuts: to restrict the angle between two outgoing muons to $\theta < 0.1$~rad, parent meson decay length to $d<2$~mm, and the number of charged tracks $N_{ch}$ with momentum $p>300$~MeV to exactly 2.}
\label{fig:signal_and_bg_spectra}
\end{figure}

The effects of our proposed cuts on the DIS dimuon background are shown in \cref{fig:signal_and_bg_spectra}. We find only $\mathcal{O}(1)$ prompt dimuon background events remain after cuts, while notably, the trident signal is very mildly affected.

\medskip
\textbf{Charged current (CC) single pion production (SPP)} Muon (anti-)neutrino CC interactions producing a single charged pion and a muon in the final state can mimic neutrino trident events if the charged hadron is mis-reconstructed as a muon. While most high-energy neutrino DIS interactions produce multiple charged tracks, as discussed above, the SPP event rate is non-negligible. We estimate these backgrounds with \texttt{GENIE v3.4.2}~\cite{Andreopoulos:2009rq,Andreopoulos:2015wxa,GENIE:2021npt} as follows. The high-energy ($>100~\textrm{GeV}$) cross section is calculated using the CSMS11 approach~\cite{Cooper-Sarkar:2011jtt} as implemented in the \texttt{GHE19\_00b} comprehensive model configuration obtained within the \texttt{HEDIS} package~\cite{Garcia:2020jwr}. For lower energies ($<100~\textrm{GeV}$), we employ the \texttt{G18\_10a\_02\_11b} tune~\cite{GENIE:2021zuu, GENIE:2021wox} (the “\texttt{G18\_10a}” model set embeds the best theoretical modeling elements implemented in GENIE~\cite{GENIE3_manual}).

At high energies relevant for LHC neutrinos, the most important contribution to the SPP in $\nu_\mu$ and $\bar{\nu}_\mu$ interactions comes from CCDIS events that, occasionally, produce only neutral particles and soft charged tracks other than the dominant $\mu^\pm\pi^\mp$ pair. On top of this, CC resonant pion production (CCRES) processes can also yield a similar signature for lower momentum exchange with the nucleus. This is when an outgoing nucleon is a neutron or proton with small kinetic energy, taking into account final-state interactions (FSI) inside the nucleus. In addition, despite giving a subdominant contribution to the predicted neutrino scattering rate, coherent pion production (COHPION), e.g., $N \nu_\mu \to N \mu^- \pi^+$, where the nucleus stays intact, also contributes to SPP backgrounds after cuts. We find that the contribution from quasi-elastic scatterings (CCQE) is quite suppressed and does not play a substantial role in our analysis. 

\begin{table}
\centering
\begin{tabular}{c|c|c|c}
\hline
\hline
\textbf{SPP} & \multicolumn{3}{c}{Number of events after cuts}\\
Process & Multiplicity & Multiplicity & All\\
 & & \& angular & (w/o final $\mu$ ID) \\
\hline 
CCDIS & $2519$ & $1362$ & $25.8$\\
CCRES & $148.1$ & $1.6$ & $0.025$ \\
COHPION & $121.9$ & $13.4$ & $0.21$ \\
CCQE & $1.8$ & $0.3$ & $0.004$ \\
\hline
\hline
\multicolumn{2}{c}{} & \textbf{Total} & $26.0^{+14.5}_{-14.1}$
\end{tabular}
\caption{Predicted single pion production background rates in FASER$\nu$2, originating from different types of neutrino scattering processes after the multiplicity, angular, and energy-loss cuts are considered. The predicted total number of neutrino scatterings is $983$k for the dominant CCDIS process, $3014$ for CCRES, $1160$ for CCQE, and $121.9$ for COHPION. We note, however, that the trident events will not be searched for among all the neutrino-induced vertices in the emulsion detector. Instead, candidate events can first be identified based on the reverse tracking of two muons. \textit{The remaining SPP backgrounds given in the $4$th column of the Table can be rejected in the analysis based on the muon ID detector placed downstream of the FASER2 spectrometer (Fig.~\ref{fig:schematic}), as discussed in the text.}
}
\label{tab:SPP}
\end{table}

\cref{tab:SPP} shows the expected background event numbers for the various SPP contributions discussed above and the impact of the cuts on the numbers. In particular, the multiplicity cut alone allows one to reduce the number of expected CCDIS-induced SPP events to a few thousand, in comparison with the total event rate of $\mathcal{O}(1\textrm{M})$. The relevant numbers for expected CCRES and COHPION events are of $\mathcal{O}(100)$. In particular, the resonant pion production rate is suppressed to a level similar to that of coherent production after considering the charged-track multiplicity. The additional angular cut suppresses even more strongly events in which soft pions are produced via nuclear resonances. 

\begin{figure}
\centering
\includegraphics[width=0.4\textwidth]{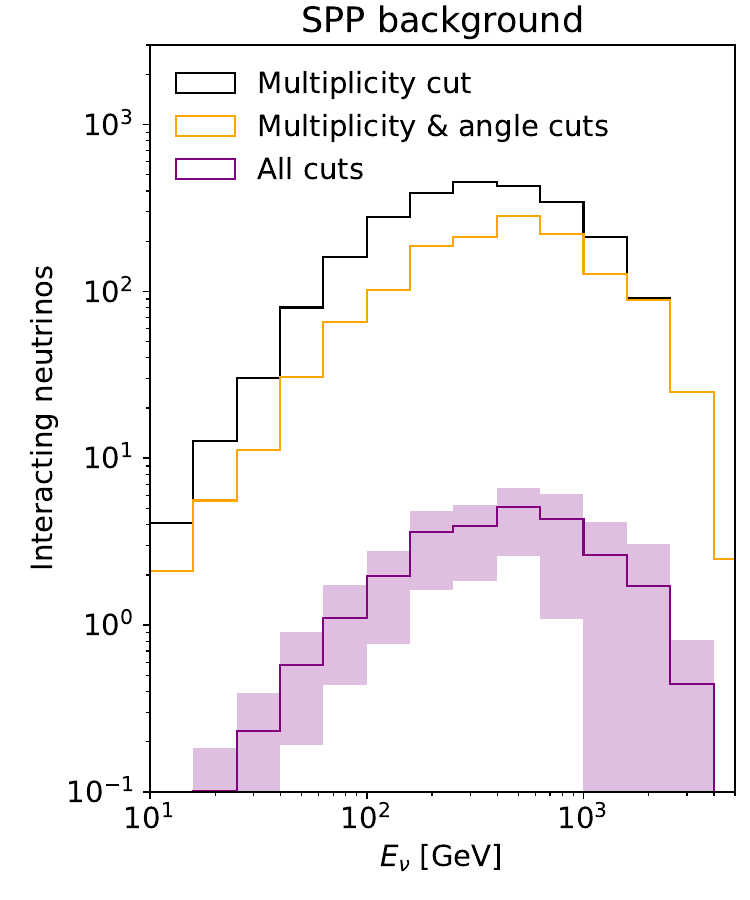}
\caption{The effect of various cuts on the single pion production background. This can be rejected based on muon vs. pion identification; see text for details.}
\label{fig:single_pi_bg}
\end{figure}

Further rejection capabilities come from considering the pion energy loss. Given their characteristic decay length, $\gamma\beta c\tau_{\pi^\pm}\simeq 7.8~\textrm{m} \times (E_\pi/m_\pi)$ and the hadronic interaction length in tungsten, $\lambda_{\textrm{had}}  = 9.946~\textrm{cm}$~\cite{ParticleDataGroup:2022pth}, boosted $\pi^\pm$s produced in high-energy neutrino interactions inside FASER$\nu$2 typically interact in tungsten layers before decaying to a muon or leaving the detector. As a result, residual backgrounds from charged pions decaying promptly are suppressed. Hence, charged pions produced in tungsten away from the downstream interface tracker are not likely to mimic long muon-like tracks in the emulsion. We estimate the total SPP background event rate after applying all the cuts above to be $N_{\textrm{BG,SPP}}\simeq 26.0^{+14.5}_{-14.1}$. We present the expected energy spectrum of these events in \cref{fig:single_pi_bg}.

The reported error bar on the SPP background event rate comes mainly from the estimated uncertainty of our Monte Carlo simulations of CCDIS events satisfying all the cuts. On top of the parton distribution function (PDF) uncertainty of a few percent on the cross section prediction ~\cite{Garcia:2020jwr}, these estimates are additionally sensitive to hadronization and FSI modeling. Due to varying predictions of different interaction models, the subdominant CCRES and COHPION event rates at high energies are subject to further uncertainties (see Refs.~\cite{Kabirnezhad:2017jmf,GENIE:2024ufm,Yan:2024kkg} for recent discussions). In particular, the employed \texttt{GENIE} implementation is based on the Berger-Sehgal model~\cite{Berger:2007rq,Berger:2008xs}; as its tuning is based on low-energy neutrino scattering data, we assume a larger ($100\%$) uncertainty in modeling these in our study.

An estimate of the uncertainty in modeling SPP can be obtained by examining results from different neutrino generators, as was done in a recent comparison of \texttt{GENIE} and \texttt{NuWro}~\cite{Juszczak:2005zs,Golan:2012wx} 
with data from MINER$\nu$A at $E_\nu \sim 3~\textrm{GeV}$~\cite{Pradhan:2024vtv}. As discussed therein, while these differences can be as large as a few tens of percent, they become much smaller at high momenta in the DIS-dominated regime. Measuring SPP events remains challenging at the high energies of our interest, so there is less available data to tune the generators. However this may be overcome with ongoing and future neutrino measurements at the LHC which would enable a more detailed study.

Importantly, even though \textsl{a priori} the estimated SPP background rate after cuts is comparable to the trident signal rate, we stress that these backgrounds can be rejected based on the final-state muon identification. This is based on the additional muon ID system in FASER2, as discussed in \cref{sec:FASERnu2}. Therefore, we do not treat these backgrounds as irreducible in further analysis. A similar conclusion holds for background events associated with kaon production that are suppressed with respect to SPP, cf. discussion in Ref.~\cite{Formaggio:2012cpf}.

\subsection{Muon-induced backgrounds\label{sec:muoninducedBG}}

Although magnets deflect most of the muons produced in $pp$ collisions at the ATLAS interaction point, a significant number of them can still reach the FASER$\nu$2 detector~\cite{Feng:2022inv}. Since the emulsion detectors collect information about events integrated over time, multiple muon tracks crossing the emulsion need to be rejected in our analysis. This can be done because they do not produce neutral vertices, i.e., vertices with no incoming track.

Muons passing through FASER$\nu$2 will produce secondary particles that subsequently interact in tungsten. In particular, one expects about $10\%$ of high-energy muons to emit photons via bremsstrahlung~\cite{Batell:2021blf}. Photons leave no tracks in the emulsion and travel about $(9/7)\,\lambda_{\textrm{rad,W}} \simeq 0.45~\textrm{cm}$ before producing an $e^+e^-$ pair, where $\lambda_{\textrm{rad,W}}$ is the radiation length in tungsten. This distance corresponds to many tungsten layers in FASER$\nu$2. The resulting (neutral) photon interaction vertex is then distant from the parent muon kink and will be reconstructed separately; see, however, Ref.~\cite{Ariga:2023fjg} for discussion about strategies to associate both vertices in the analysis.

Occasionally, high-energy photons can also produce dimuon pairs, $\gamma N \to N \mu^+\mu^-$, mimicking the neutrino trident signal. However, the relevant production rate is suppressed by the lepton mass ratio, $(m_e/m_\mu)^2\sim 2\times 10^{-5}$. Given the total muon rate predicted for FASER$\nu$2~\cite{sabate-gilarte:ipac2023-mopl018}, we expect about $\mathcal{O}(10^5)$ such events in the detector. The bremsstrahlung photons are typically of GeV energy, resulting in soft muons that are rejected based on their energy loss in tungsten. Only the high-energy tail of the bremsstrahlung photon-induced dimuon spectrum can mimic neutrino tridents, resulting in an additional 2 orders of magnitude or so suppression.

Since the number of such background events might still seem substantial even after all these cuts, we stress that they can in any case be rejected by vetoing incident muons that enter the emulsion detector and associating this with time-coincident dimuon signals in the interface tracker and spectrometer. Notably, the muon veto in the current FASER experiment is extremely efficient~\cite{FASER:2023tle}. Given the higher statistics of such background events, they could also be used to test and calibrate the reverse tracking procedure. Finally, neutrino-induced muons produced inside FASER$\nu$2 could contribute to photon bremsstrahlung backgrounds. However, the much lower expected statistics of such events, renders these backgrounds negligible. 

We conclude that, although they do need to be taken into account in the analysis, muon-induced backgrounds can be efficiently rejected in the neutrino trident measurement in FASER$\nu$2.

\subsection{Statistical significance and uncertainty} 

Using the above estimates for the total number of signal ($s=18.6$ after cuts) and background ($b = 0.55$) events in FASER$\nu$2, we estimate the expected statistical significance of the dimuon trident discovery in FASER$\nu$2 to be 
$9.9\sigma$ for the baseline neutrino flux used in our study. Similarly, we estimate $11.1\sigma$ ($7.3\sigma$) significance for the most optimistic (pessimistic) predictions shown in \cref{fig:trident_generators_contributions}. We reiterate that while the flux uncertainties remain substantial at this stage, they will be significantly reduced to even sub-percent level with the CCDIS data collected in the current FASER and SND@LHC experiments and the future FASER$\nu$2 detector~\cite{Kling:2023tgr}. The proposed FASER$\nu$2 detector thus provides a very promising tool for the first precision measurement of dimuon neutrino tridents.

We note that besides the efficiency of the cuts employed in our study, additional detector-level efficiencies will affect the trident signal rates in FASER$\nu$2. We find that these need to be higher than $40\%$ to guarantee the $>5\sigma$ discovery potential. Last but not least, the trident search in FASER$\nu$2 relies on excellent background rejection potential. In particular, if the residual SPP background mentioned in \cref{tab:SPP} ($b\simeq 26$) cannot be rejected based on the additional iron block and muon ID, cf. discussion in \cref{sec:FASERnu2}, this will translate into a reduced statistical significance of the trident measurement to about $3.3\sigma$. The SPP and other backgrounds must be suppressed down to $N_{\textrm{BG}}\lesssim 9$, in order to guarantee discovery at $5\sigma$ significance. Thus our study highlights certain detector requirements that would have to be considered for the future HL-LHC period in order to allow for trident measurement. As long as the assumed background rejection and signal detection efficiencies can be achieved, the expected accuracy of the dimuon trident production rate in FASER$\nu$2 will be of order $23\%$, and the statistical uncertainty on the number of signal events will dominate.

\section{Constraining four-fermi muon interactions}
\label{sec:BSM}

As the typical momentum transfer in trident events is significantly below the electroweak scale, weak interactions can be described in terms of the effective interaction~\cite{Altmannshofer:2019zhy}
\begin{align}
\mathcal{H}_{\rm eff} =
\frac{G_F}{\sqrt{2}} \bigg(
  &g_V^{ijkl} (\bar{\nu}_i \gamma_\alpha P_L \nu_j)(\bar{\ell_k} \gamma^\alpha \ell_l)\nonumber\\
+ &g_A^{ijkl} (\bar{\nu}_i \gamma_\alpha P_L \nu_j)(\bar{\ell_k} \gamma^\alpha \gamma_5 \ell_l) \bigg) ~,
\end{align}
where $G_F$ is the Fermi coupling, $P_L$ the left-handed projection, $\gamma_\alpha$ the Dirac gamma matrices and $g_V^{ijkl}$ ($g_A^{ijkl}$) the (axial-)vector coupling with $i,j,k,l \in \{e,\mu,\tau\}$. 
However, the couplings involving electrons are stringently constrained by bounds from LEP~\cite{2013119}, and the incoming neutrinos expected in forward neutrino detectors at the LHC are predominantly muon neutrinos. So we focus on possible modifications to the vector and axial-vector couplings $g_V^{\mu\mu\mu\mu} \equiv g_V$ and $g_A^{\mu\mu\mu\mu} \equiv g_A$, respectively, and parametrize new physics interfering with muonic interactions as
\begin{equation}
\begin{cases}
g_V = 1 + 4\sin^2\theta_\mathrm{W} + \Delta g_V\\
g_A = -1 + \Delta g_A,
\end{cases}
\end{equation}
where $\theta_\mathrm{W}$ is the Weinberg angle. 
A possible UV completion of such a theory is provided for example by a heavy muonphilic $Z'$~\cite{Greljo:2021npi}.
For a fixed incoming neutrino energy $E_\nu$, the resulting cross section has approximately the form of an elliptical paraboloid in $(\Delta g_V, \Delta g_A)$ space, 
with the minimum at $(\Delta g_V, \Delta g_A) = (-1 - 4\sin^2\theta_\mathrm{W}, 1)$ lying below the SM value at $(\Delta g_V, \Delta g_A) = (0,0)$. 

\begin{figure*}
    \centering
    \includegraphics[width=0.47\textwidth]{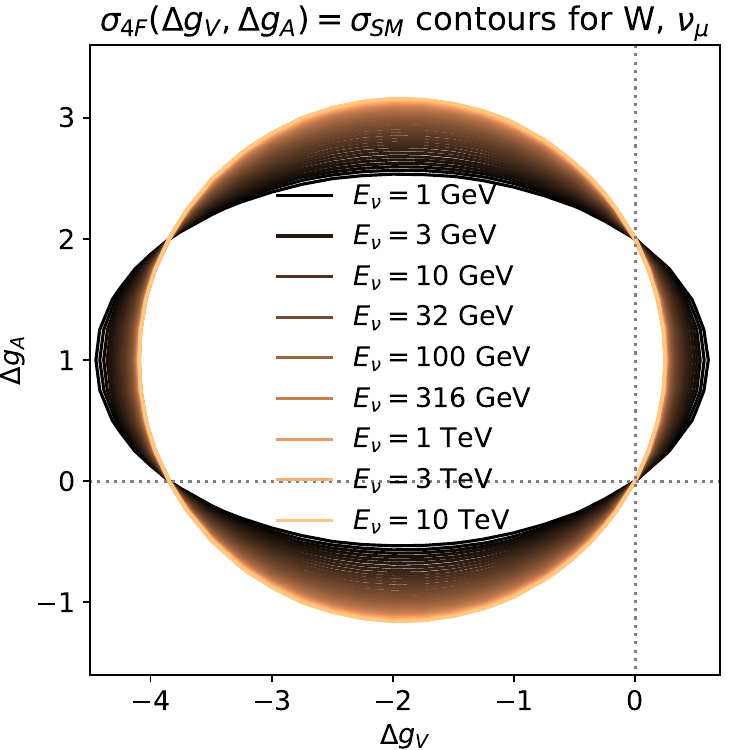}
    \includegraphics[width=0.47\textwidth]{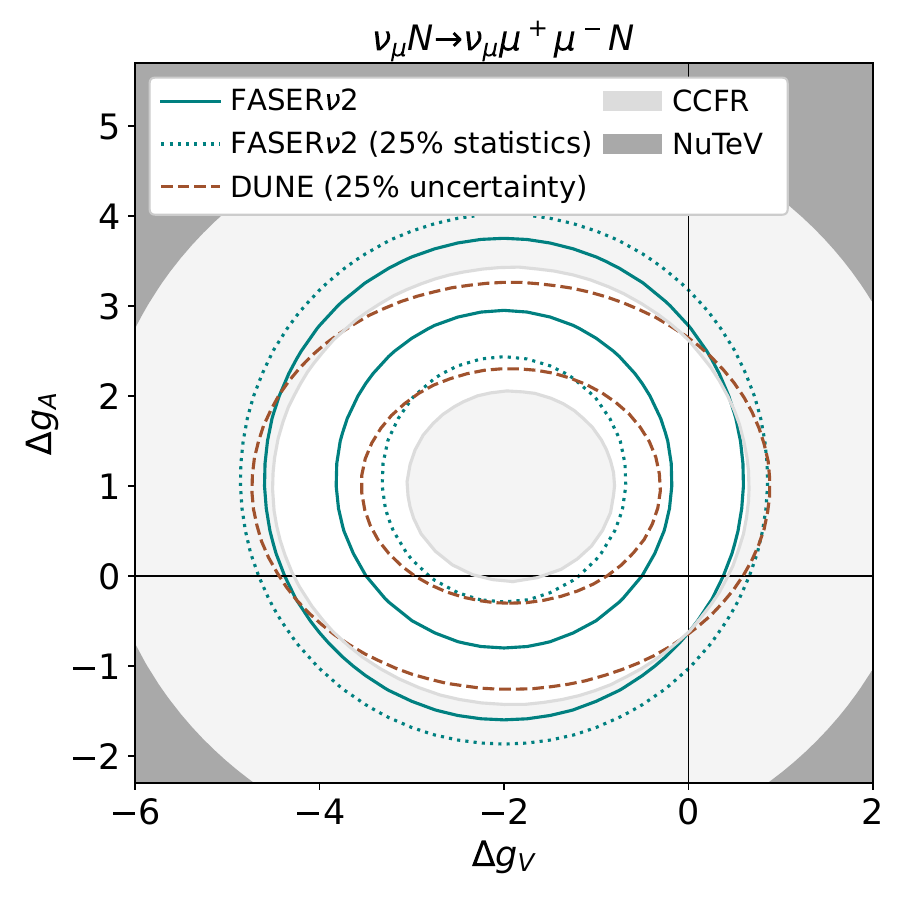}
    \caption{\textit{Left:} Contours of modifications to axial and vector couplings $\Delta g_A$ and $\Delta g_V$, which result in cross sections identical to the SM values in muon neutrino interactions with tungsten. The intersection of the dotted lines indicates the SM. The dark (bright) contours correspond to low (high) incoming $E_\nu$, with the legend illustrating the color gradient for a selection of $E_\nu$ values. Since the observed event number is directly proportional to the cross section, this illustrates that differently-shaped constraints are expected from experiments performed at different energies. 
    \textit{Right:} Modifications to axial and vector couplings excluded by existing CCFR and NuTeV measurements (gray), compared to the projected bounds from DUNE assuming a 25\% uncertainty cross section measurement (dashed brown), and at FASER$\nu$2 for 100\% (solid teal) or 25\% (dotted teal) of the expected data. The solid black lines intersect at $(\Delta g_V,\Delta g_A)=(0,0)$, corresponding to the SM. The NuTeV bound is computed following Ref.~\cite{Altmannshofer:2019zhy} and assumes equal BSM to SM cross section ratios at CCFR and NuTeV due to their similar energies. All sensitivities are at 95\% confidence level. 
    }
    \label{fig:FFconstraint}
\end{figure*}

We note that the coherent $N \nu_\mu \to N \nu_\mu \mu^+ \mu^-$ scattering cross section~\cite{Lovseth:1971vv, Brown:1972vne} (see also~\cite{GE2017164, Ballett:2018uuc, Altmannshofer:2019zhy})
depends on the leptonic tensor 
$L_{\alpha\beta} = \sum_{\rm spins} A_\alpha A_\beta^\dagger$,
where
\begin{align}
A_\alpha
=
(&\overline{u}' \gamma_\mu P_L u ) \nonumber\\
\times
\bigg(\overline{u}_-
      &\bigg[ \gamma_\alpha
              \frac{\slashed{p}_- - \slashed{q} + m}{(p_- - q)^2 - m^2}
              \gamma^\mu (g_V + g_A \gamma_5)
              \nonumber \\
            -&\gamma^\mu (g_V + g_A \gamma_5)
              \frac{\slashed{p}_+ - \slashed{q} + m}{(p_+ - q)^2 - m^2}
              \gamma_\alpha
      \bigg]
      v_+
\bigg).
\label{LeptonicA}
\end{align}
Since the terms proportional to the final state charged lepton momenta $\slashed{p}_\pm$ and momentum transfer $\slashed{q}$ carry more gamma matrices than those proportional to the muon mass $m$, the $\gamma_5$ matrix decreases the growth of the cross section as a function of $|g_A|$. In contrast, the $g_V$ direction remains unaffected. However, the cross section becomes symmetric at high $E_\nu$ as incoming neutrino momentum is transferred mainly to a charged lepton, making the $m$-dependent terms negligible. This effect is illustrated by the SM-equivalent contours of the cross section in Fig.~\ref{fig:FFconstraint} (left), shown together with the projected exclusion bounds for experiments at various energies (right), each displaying a distinct ellipticity. It should be noted that the exploitation of this effect does not require a particular $E_\nu$ binning, but is determined by the energy scale of the bulk of the interactions. Since the $E_\nu$ resolution of the trident signal is limited by missing energy due to the final state neutrino, we proceed conservatively by assuming that the total number of observed events across the entire $E_\nu$ range of the experiment is Poisson distributed. The constraints are then found as the contours corresponding to the 95\% confidence level deviations up and down from the SM expectation, using an interpolation grid as a function of $\Delta g_V$ and $\Delta g_A$.

Although trident event statistics are limited at the FPF, our results indicate that FASER$\nu$2 would complement the existing limits on $\Delta g_V,\Delta g_A$ obtained in Ref.~\cite{Altmannshofer:2019zhy} from CCFR cross section measurements~\cite{PhysRevLett.75.3993}, cf. also Refs.~\cite{Ismail:2020yqc,Falkowski:2021bkq,Kling:2023tgr} for FPF bounds on effective interactions between neutrinos and quarks. 
For reference, we also indicate the constraints obtained from the NuTeV measurement~\cite{NuTeV:1999wlw} assuming a BSM to SM cross section ratio the same as that for CCFR, given the similar energies of the two experiments.
Moreover, already with only 25\% of the expected data, FASER$\nu$2 can be expected to partially enhance the projected bounds on $\Delta g_A$ obtained for DUNE with $E_\nu = \mathcal{O}(1 {\rm GeV})$~\cite{Altmannshofer:2019zhy}.
This is due to the cross section dependence on $E_\nu$, which reaches values of $\mathcal{O}(1 {\rm TeV})$ at FASER$\nu$2, rendering the FASER$\nu$2 bounds more symmetric in $\Delta g_V, \Delta g_A$ than the low-energy experiments.

\section{Conclusions}
\label{sec:conclusions}

 The study of neutrino interactions has long been a key tool for discoveries in particle physics, and a new opportunity has now opened up with dedicated forward detectors to study the naturally collimated high-energy neutrinos emanating from the LHC collision points. In this study, we have analyzed the prospects for measuring neutrino trident events, finding that in the coming High-Luminosity data-taking period, an $\mathcal{O}(10~\textrm{ton})$ detector placed in the forward kinematical region of the LHC could measure tens of dimuon trident events. We have discussed in detail the trident signature and possible background rejection strategies for the proposed FASER$\nu$2 experiment. We estimate that with this detector a statistical significance between $7.3$ and $11.1\sigma$ can be achieved for the first unambiguous measurement of the neutrino trident process, by adopting a reverse tracking strategy to identify signal and reject backgrounds in the analysis. 

Besides testing predictions of the SM, the dimuon neutrino trident measurements in FASER$\nu$2 can also be used to probe new physics. We have analyzed prospects for constraining BSM contributions to effective four-Fermi contact operators between muon neutrinos and muons and highlighted the complementarity between low- and high-energy experiments setting such bounds.

We have also provided estimates for other leptonic final states. In particular, event statistics of neutrino trident interactions producing $e^+e^-$ or $e^\pm\mu^\mp$ are similar or larger than dimuon event rates, although they are more difficult to disentangle from backgrounds. In addition, of $\mathcal{O}(10)$ events with the $\mu^\pm\tau^\mp$ final states are expected in some of the proposed detectors, which could enable the first-ever observation of such trident final states. This, however, would require a modified detection strategy as well as background mitigation strategies that differ from our discussion and need to be studied separately.

The neutrino trident measurement at the LHC would also pave the way for similar studies in future colliders. In particular, a collimated beam of high-energy neutrinos should also be produced in the forward region of the Future Circular Collider operating with proton beams at $100~\textrm{TeV}$ center-of-mass energy (FCC-hh)~\cite{FCC:2018byv, FCC:2018vvp}. Since the corresponding neutrino interaction rates will exceed those at the LHC predictions by about two orders of magnitude~\cite{MammenAbraham:2024gun}, rare neutrino scattering processes can then be measured with unprecedented precision. Increased neutrino energies could additionally allow for studying on-shell $W$-boson production~\cite{Zhou:2019vxt} in a laboratory setup.

Despite decades of experimental efforts, neutrinos still hold many secrets. Measuring high-energy neutrino interactions at hadron colliders would provide a new window to unravel them, as was anticipated when the LHC was first proposed \cite{DeRujula:1984pg}. The neutrino physics program ought thus to be an essential part of all planning for future hadron colliders.

\section*{Acknowledgements}

We thank Felix Kling for valuable comments on the manuscript, and Akitaka Ariga, Tomoko Ariga, Kingman Cheung, Milind Diwan, Minoo Kabirnezhad, C.J. Ouseph, Marta Sabat$\acute{\textrm{e}}$-Gilarte, Sourav Sarkar, and Jan Sobczyk for helpful discussions. We are grateful to Max Fieg and Felix Kling for sharing event files. TM and ST are supported by the National Science Centre, Poland (research grant No. 2021/42/E/ST2/00031). TM is also supported in part by U.S.~National Science Foundation Grants PHY-2111427 and PHY-2210283 and Heising-Simons Foundation Grant 2020-1840. ST was also partially supported by the grant ``AstroCeNT: Particle Astrophysics Science and Technology Centre" carried out within the International Research Agendas programme of the Foundation for Polish Science, financed by the European Union under the European Regional Development Fund. ST was additionally partly supported by the European Union’s Horizon 2020 research and innovation program under grant agreement No 952480 (DarkWave). KX is supported by U.S. National Science Foundation (Grant Nos. PHY-2013791 and PHY-2310497). The research of W.A. is supported by the U.S. Department of Energy (Grant No. DE-SC0010107). SS is supported by the UK STFC. B.Z. is supported by the Fermi Research Alliance, LLC, acting under Contract No.\ DE-AC02-07CH11359.

\onecolumngrid
\appendix
\onecolumngrid
\section{Form factors}
\label{sec:formfactors}

The computation of trident cross sections follows the procedure outlined in Ref.~\cite{Altmannshofer:2019zhy}. 
The calculation relies on form factors given by
\begin{equation}
F_N(q^2) = \int dr r^2 \frac{\sin(qr)}{qr} \rho_N(r).
\label{formfactor}
\end{equation}
For argon, the spherically symmetric charge density distribution $\rho_N(r)$ is expressed in terms of the spherical Bessel function of zeroth order as
\begin{equation}
\rho_N(r) = 
\begin{cases}
\mathcal{N} \sum_n a_N^{(n)} j_0(n \pi r/R_n) &\textrm{, for } r < R_N,\\
0 &\textrm{, for } r > R_N,
\end{cases}
\label{rho_Fourier-Bessel}
\end{equation}
where $a_N$ and $R_N$ are the Fourier-Bessel coefficients listed in Ref.~\cite{DEVRIES1987495}.
As these coefficients are however unavailable for tungsten, the functional form
\begin{equation}
\rho_N(r) = \frac{\mathcal{N}}{1 + \exp((r-r_0)/\sigma)},
\label{rho_approx}
\end{equation}
is employed, with $r_0 = 1.18 \textrm{ fm} \times A^{1/3} - 0.48 \textrm{ fm}$ and $\sigma = 0.55 \textrm{ fm}$ --- found to best correspond to the result of Eq.~\eqref{rho_Fourier-Bessel} in the case of argon and iron. The normalization factor $\mathcal{N}$ in Eqs.\eqref{rho_Fourier-Bessel}--\eqref{rho_approx} is determined by requiring
\begin{equation}
\int dr r^2 \rho_N(r) = 1 
\end{equation}
yielding $F_N(0) = 1$.

\begin{figure}[H]
\centering
\includegraphics[width=0.45\textwidth]{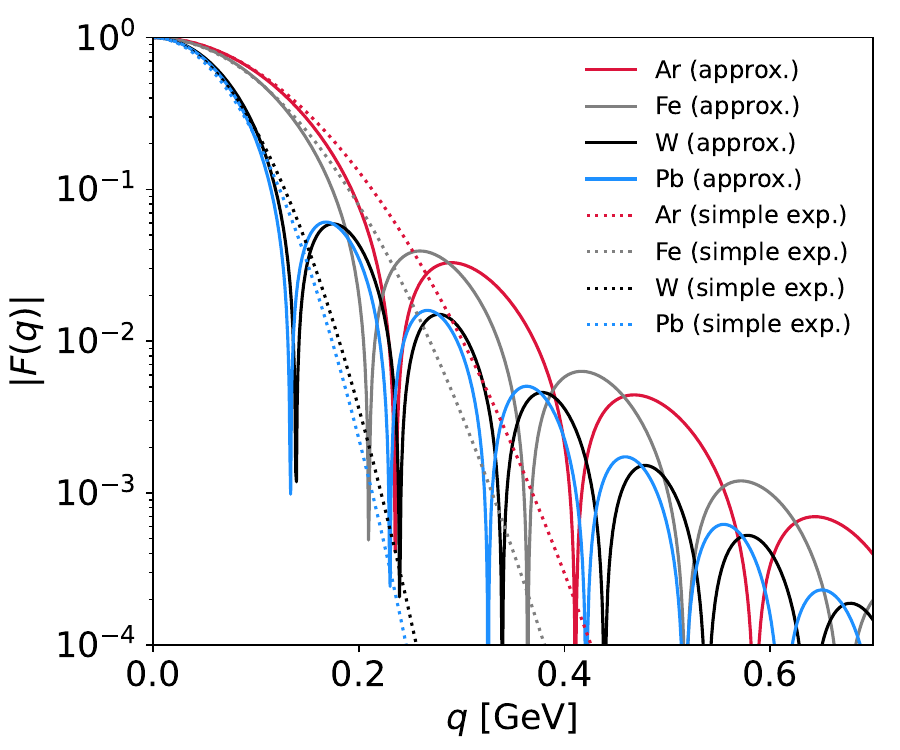}
\includegraphics[width=0.45\textwidth]{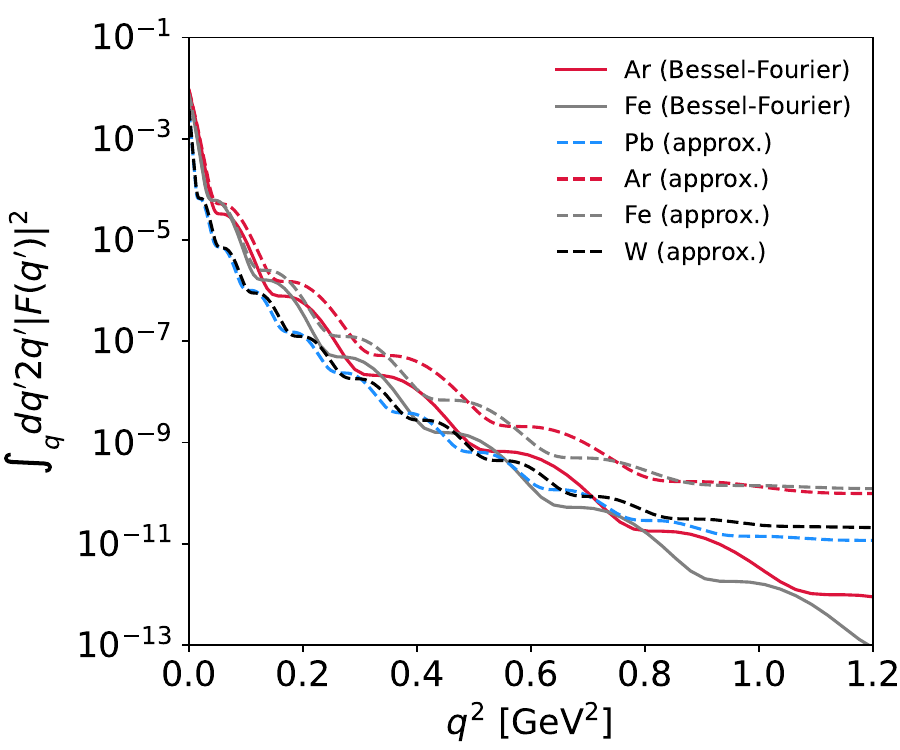}
\caption{
\textit{Left:} The approximate form factors for argon, iron, tungsten, and lead using Eq.~\eqref{rho_approx}, compared with the simple exponential form 
$F_N(q) = e^{- a^2 q^2/10}$,
$a = A^{1/3} 1.3 $~fm. 
\textit{Right:} Integrals of the approximate form factors squared and the interpolation grids provided in the original computation of Ref.~\cite{Altmannshofer:2019zhy} as in Eq.~\eqref{rho_Fourier-Bessel}, for comparison with our approximations for argon and iron.
}
\label{formFactors}
\end{figure}

The approximate form factors obtained using Eq.~\eqref{rho_approx} are depicted in the left panel of Fig.~\ref{formFactors}, together with the simple exponential form 
$F_N(q) = e^{- a^2 q^2 / 10}$
with
$a = A^{1/3} 1.3 $~fm. 
To illustrate the level of agreement between Eqs.~\eqref{rho_Fourier-Bessel}--\eqref{rho_approx}, the right panel of Fig.~\ref{formFactors} shows a comparison of the integrals of the approximated form factors for argon and iron to those obtained using Eq.~\eqref{rho_Fourier-Bessel}, as well as the integral of the approximate form factor for tungsten. These integrals enter the trident cross section computation in the form of interpolation grids.

\clearpage
\bibliography{references}

\end{document}